# Spontaneous Activity in the Visual Cortex is Organized by Visual Streams

Kun-Han Lu[2,3], Jun Young Jeong[2], Haiguang Wen[2,3], Zhongming Liu[*,1,2,3]

[1]Weldon School of Biomedical Engineering

[2]School of Electrical and Computer Engineering

[3]Purdue Institute for Integrative Neuroscience

Purdue University, West Lafayette, IN, USA

**\*Correspondence**

Zhongming Liu, PhD

Assistant Professor of Biomedical Engineering

Assistant Professor of Electrical and Computer Engineering

College of Engineering, Purdue University

206 S. Martin Jischke Dr.

West Lafayette, IN 47907, USA

Phone: +1 765 496 1872

Fax: +1 765 496 1459

Email: zmliu@purdue.edu

**Abstract**

Large-scale functional networks have been extensively studied using resting state functional magnetic resonance imaging. However, the pattern, organization, and function of fine-scale network activity remain largely unknown. Here we characterized the spontaneously emerging visual cortical activity by applying independent component analysis to resting state fMRI signals exclusively within the visual cortex. In this sub-system scale, we observed about 50 spatially independent components that were reproducible within and across subjects, and analyzed their spatial patterns and temporal relationships to reveal the intrinsic parcellation and organization of the visual cortex. We found that the visual cortical parcels were aligned with the steepest gradient of cortical myelination, and organized into functional modules segregated along the dorsal/ventral pathways and foveal/peripheral early visual areas. In contrast, cortical retinotopy, folding, and cytoarchitecture impose limited constraints to the organization of resting state activity. From these findings, we conclude that spontaneous activity patterns in the visual cortex are primarily organized by visual streams, likely reflecting feedback network interactions.

**Introduction**

Resting state functional magnetic resonance imaging (rs-fMRI) has been widely explored to map resting state networks (RSNs) that collectively report the brain's intrinsic functional organization (Fox and Raichle, 2007). These networks consist of temporally correlated regions (Biswal et al., 1995; Van Dijk et al., 2010), arise from structural connections (Honey et al., 2007; Van Den Heuvel et al., 2009), resemble and predict task activations (Smith et al., 2009; Cole et al., 2014; Tavor et al., 2016), and distinguish individual subjects (Finn et al., 2015) or diseases (Fox and Greicius, 2010). Although brain activity spans a variety of spatial scales (Yoshimura et al., 2005; Doucet et al., 2011; Hutchison et al., 2013), RSNs have been mostly characterized in the whole brain, where their functions are empirically and coarsely defined in terms of motor (Biswal et al., 1995), vision (Yeo et al., 2011), default-mode (Greicius et al., 2003), attention (Fox et al., 2006), salience (Seeley et al., 2007), and so on. In finer spatial scales, patterns of spontaneous activity and connectivity remain unclear, but may bear more specific functional roles related to perception, behavior, or cognition (Kenet et al., 2003; Yoshimura et al., 2005; Wang et al., 2013; Long et al., 2014; Wilf et al., 2015; Lewis et al., 2016).

In this regard, the visual cortex is a rich and ideal benchmark. It has been characterized in terms of cellular architectonics (Amunts et al., 2000; Glasser and Van Essen, 2011), structural connections (Felleman and Van Essen, 1991; Salin and Bullier, 1995), cortical folding (Fischl et al., 2008; Benson et al., 2012), functional pathways (Hubel and Wiesel, 1962; Ungerleider and Haxby, 1994), and neural coding (Ohiorhenuan et al., 2010; Guclu and van Gerven, 2015). In recent studies, correlation patterns of spontaneous activity in the visual cortex have been examined and compared with retinotopy (Heinzle et al., 2011; Yeo et al., 2011; Jo et al., 2012; Butt et al., 2013; de Zwart et al., 2013; Gravel et al., 2014; Raemaekers et al., 2014; Arcaro et al., 2015; Bock et al., 2015; Striem-Amit et al., 2015; Wilf et al., 2015; Dawson et al., 2016; Genc et al., 2016; Lewis et al., 2016). Since much of the visual cortex has visual-field maps – topographic representations of the polar angle and eccentricity (Wandell et al., 2007), it is reasonable to initially hypothesize that resting state activity is also intrinsically organized by retinotopy.

Support for this hypothesis comes in part from the finding that in early visual areas (e.g. V1-V3), correlations in spontaneous activity are generally higher between locations with similar eccentricity representations (Heinzle et al., 2011; Yeo et al., 2011; Jo et al., 2012; Gravel et al., 2014; Arcaro et al., 2015; Striem-Amit et al., 2015; Dawson et al., 2016; Genc et al., 2016; Lewis et al., 2016). However, such eccentricity-dependent functional connectivity has rarely been reported beyond early visual areas (Baldassano et al., 2013; Striem-Amit et al., 2015); it may be relatively weak (Wilf et al., 2015) or observable only after regressing out large-scale activity (Raemaekers et al., 2014); and it may also be confounded by the decay of local connectivity over cortical distance (Butt et al., 2013; Dawson et al., 2016). Further against this hypothesis, spontaneous activity preserves little or less polar-angle dependence in intra-hemispheric correlations (Gravel et al., 2014; Bock et al., 2015; Wilf et al., 2015), and shows strong correlations between bilateral V1 locations despite their lack of common receptive fields or direct connections (Jo et al., 2012; de Zwart et al., 2013). As such, the retinotopic organization of fine-scale resting activity in the visual cortex is questionable (Butt et al., 2013; Wilf et al., 2015). The previously reported eccentricity-dependent spontaneous activity and connectivity may be attributed to alternative representations that partly overlap with the eccentricity representation in early visual areas, as opposed to retinotopy *per se*.

In fact, cortical representations of the peripheral and central visual fields partly overlap with the magnocellular and parvocellular streams (Schiller et al., 1990; Nassi and Callaway, 2009), and extend onto the dorsal and ventral pathways for visual action and perception, respectively (Goodale and Milner, 1992; Ungerleider and Haxby, 1994). Along these pathways, feedforward neuronal circuits convey and integrate not only visual positions, but also increasingly complex visual or conceptual features (Martin, 2007; Hasson et al., 2008; Yamins and DiCarlo, 2016). Top-down feedback connections are not or less retinotopically organized than are feedforward pathways (Salin and Bullier, 1995). The complex interplay between feedforward and feedback processes is essential for natural vision (Rao and Ballard, 1999; Gilbert and Li, 2013), but remains largely unclear in a stimulus-free resting state. Thus, spontaneously emerging networks in the visual cortex may not readily fit the retinotopic organization, or arguably any other presumable organizations.

What is needed is a data-driven analysis of resting-state activity in a finer spatial scale, unbiased by any presumed areal definition or organizational hypothesis. For this purpose, independent component analysis (ICA) is well suited but has not been applied to finer spatial scales, to our knowledge, despite its wide application to whole-brain fMRI signals (Damoiseaux et al., 2006). Unlike the correlation analysis (Baldassano et al., 2012; Genc et al., 2016), ICA also has the advantages of being multivariate and data-driven, thereby bypassing the potential bias from any narrowly-focused hypothesis (Calhoun et al., 2009). Here, we explored a new application of ICA for mapping cortical visual areas and networks based on rs-fMRI signals within the visual cortex. In this sub-system scale, the fine-grained activity patterns derived from ICA were systematically characterized, interpreted, and evaluated for their test-retest reproducibility and individual variations. Moreover, they were also compared against cortical folding (Destrieux et al., 2010), retinotopy (Abdollahi et al., 2014), cytoarchitecture (Van Essen et al., 2012a), myeloarchitecture (Glasser et al., 2014), and the latest multimodal cortical parcellation (Glasser et al., 2016). Clustering analysis further reveals that spontaneously emerging network patterns in the human visual cortex are not retinotopically organized; instead, they are temporally clustered into three functional modules: the dorsal pathway, the ventral pathway, and the foveal and peripheral sub-divisions of early visual areas.

**Materials and Methods**

*Subjects and Data*

We used the rs-fMRI data released from the Human Connectome Project (HCP) (Van Essen et al., 2013). Briefly, we randomly selected 201 independent healthy subjects; for each subject, we used the data from two resting state sessions (session 1 for test and session 2 for retest); each session was 14 minutes and 33 seconds with the eyes open and fixated.

As elaborated elsewhere (Van Essen et al., 2012b), data were acquired in a 3-tesla MRI system with a 32-channel head coil (Skyra, Siemens, Germany). The rs-fMRI data we used were acquired with a single-shot, multiband-accelerated, gradient-recalled echo-planar imaging with nominally 2mm isotropic spatial resolution and 0.72s temporal resolution, and left-to-right phase-encoding. In addition, structural images with $T_1$ and $T_2$-weighted contrast were both acquired with 0.7mm isotropic resolution.

As elaborated elsewhere (Glasser et al., 2013), the structural images were non-linearly registered to the Montreal Neurological Institute (MNI) template, where the images were combined and segmented to generate cortical surfaces. The fMRI images were corrected for slice timing and motion, aligned to structural images, normalized to the MNI space, projected onto the cortical surfaces, and co-registered across subjects. In addition to the minimal preprocessing described above, we removed the slow trend in the fMRI time series by regressing out a fourth-order polynomial function, and subtracted the mean and standardized the signal variance. Note that spatial smoothing was not performed to minimize spurious correlations in neighboring voxels.

*Independent Component Analysis*

We applied the ICA to the rs-fMRI signals within a mask of the visual cortex (Fig. 1.A), defined by a system-level functional parcellation of the human cortex (Yeo et al., 2011). The fMRI time series within the mask was temporally standardized and concatenated across subjects. Infomax ICA (Bell and Sejnowski, 1995) was used to decompose the concatenated rs-fMRI data into 59 spatially independent components within the visual cortex. Here, the number of independent components, 59, was determined by maximizing the Laplace approximation of the posterior probability of the ICA model order (Beckmann and Smith, 2004). The spatial pattern of each component was converted to a z-score map by dividing the IC weight at each voxel by the standard deviation of voxel-wise residual noise that could not be explained by the ICA model (Beckmann and Smith, 2004). All color scale for displaying the spatial IC maps in this study are z-scores.

*Reproducibility test*

Furthermore, we evaluated the test-retest reproducibility of the ICA results. In a group level, the ICA applied to the data in one session (i.e. session 1) was also applied to the data in a repeated session (i.e. session 2); both sessions were from the same group of subjects, and the data was concatenated in the same order across subjects. We calculated the absolute value of the spatial correlation (r) between every component from session 1 and every component from session 2, and paired these components across sessions into distinct pairs to maximize the sum of their absolute spatial correlations. More specifically, we used an iterative procedure toward the optimal pairing: we began with identifying a pair of ICs (i.e. one from session1, and the other from session2) with the highest correlation; then we paired these two ICs, and excluded them from subsequent pairing, which continued until all ICs were paired. Note that the absolute spatial correlation was used since a reproducible IC could show the same spatial distribution despite opposite polarity. A threshold ($|r| \geq 0.4$) was used to identify reproducible components for further interpretation (Fig. 1.C).

We also explored the potential confounding effects of head motion on fine-scale independent components. For each subject, we regressed out the time series of six motion correction parameters prior to group-level ICA. We compared the ICs obtained without and with the above motion correction, and further identified and excluded those ICs that appeared inconsistent solely due to this preprocessing step.

*Modularity analysis*

For all the ICs that were reproducible and unaffected by head motion correction, we computed the temporal correlations between different components. Such correlations were first calculated based on component time series from each subject, and then averaged across subjects. This procedure prevented the resulting correlations from being dominated or biased by inter-subject variations. To evaluate the statistical significance of the between-component correlation, we converted the correlation coefficient to the z-score using the Fisher's r-to-z transform separately for each subject, and then applied a one-sample t-test (with dof=200, significance level at 0.01, and Bonferroni correction) to the z-scores from all the subjects.

We also applied the Louvain modularity analysis (Blondel et al., 2008) to the cross-component correlation matrix averaged across subjects. It assigned individual ICs to different modules, such that the temporal correlations were higher within modules but lower between modules. The modularity analysis, including the determination of the number of modules, was conducted with the algorithm (a Matlab function: *modularity_louvain_und_sign*) implemented in Brain Connectivity Toolbox (Rubinov and Sporns, 2010). The modularity index, Q, which quantified the goodness of modularity partitions (ranged from 0 to 1), was obtained. To evaluate the statistical significance of Q, we randomly shuffled the values within the correlation matrix 10,000 times and computed Q for each permutation given the same module assignment. This generated a null distribution of Q, against which the p value was computed for the Q value without permutation. To visualize the modular organization on the cortical surface, we represented each IC by a sphere located at the peak location in the component map, and color-coded every IC by its module membership.

*Computation of cortical distance*

To examine the effect of cortical distance on temporal correlations between components, we first identified the peak location in each component map (hereafter referred to as the component centroid). For bilaterally-distributed components, two centroids were identified: one at each hemisphere. We computed the geodesic cortical distances along the cortical surface between different component centroids by using HCP's "-surface-geodesic-distance" function. The cortical distances were computed for each subject and for each hemisphere, and then averaged across subjects.

The relationship between the temporal correlations and the corresponding cortical distances was examined and fitted by a rational function based on least-squares estimations (the "lsqcurvefit" function in Matlab). Separately for each hemisphere, we defined the spatial affinity between components as the reciprocal of the cortical distances between the corresponding centroids. Then we applied the Louvain modularity analysis to the spatial-affinity matrix in the same way as for the modularity analysis of functional connectivity. We further compared such anatomically-defined modules with the functional modules obtained on the basis of the temporal correlations between components.

*Intrinsic functional parcellation of the visual cortex*

From all the ICs, we created a group-level intrinsic functional parcellation of the visual cortex with an increasing level of granularity. Specifically, we defined a feature vector for each voxel in the visual cortex. This feature entails the weights by which the individual time series of different ICs were linearly combined to explain the fMRI signal observed at each voxel. Then we grouped the cortical voxels into distinct parcels by applying the k-means clustering to the corresponding feature vectors using a correlation-based "distance" and 1,000 replications with random initialization. The number of clusters (k) was empirically set to 10, 20, 30, 40. The parcels were color-coded from 0 to 1 in an ascending order according to the averaged "distance" within each parcel. We preferred this k-means clustering analysis to a "winner-take-all" alternative, in which each voxel was assigned to only one IC with the greatest weight (among all the ICs) at the given voxel. This was because single voxel time series were not necessarily represented by only one IC, but instead often by a few ICs.

To facilitate interpretation, the ICA components and the parcellation derived from them were compared against conventional visual areas or cortical parcellation based on various structural and/or functional properties, including myeloarchitecture (Glasser et al., 2014), cytoarchitecture (Eickhoff et al., 2005), cortical folding (Destrieux et al., 2010), retinotopic mapping (Abdollahi et al., 2014), and multimodal parcellation (Glasser et al., 2016).

*Dual-regression and individual-level parcellation*

Following group ICA, we also used dual regression (Filippini et al. 2009) against each subject's fMRI data to characterize subject-specific ICA maps (Tavor et al. 2016). Briefly, we first applied multiple regression to the spatial domain, using the group-level ICA spatial maps as a set of spatial regressors to obtain individual time series that was associated with each group-level spatial map based on the subject-specific fMRI data; after normalizing these individual-level time series to a zero mean and a unitary variance, we applied multiple regression to the time domain, by using the normalized individual time series as temporal regressors to obtain the subject-specific ICA spatial maps. From the individual-level ICA maps, we also used the k-means clustering analysis as mentioned above to create subject-specific parcellation of the visual cortex, and compared it against the group-level parcellation.

**Results**

*Intrinsic activity patterns within the visual cortex*

Here, we explored a data-driven analysis of spontaneous activity confined to the human visual cortex. In this finer sub-system scale, our goal was to characterize and map intrinsic activity patterns in order to delineate cortical visual areas and networks independent of any task context or any presumed organizational hypothesis. Towards this goal, we used a cortical mask (Fig. 1.A), based on a systems-level parcellation of the entire cortex (Yeo et al., 2011), to only select rs-fMRI activity within the human visual cortex for group ICA. The selected rs-fMRI data from 201 healthy human subjects in the Human Connectome Project (HCP) were temporally standardized within each subject and concatenated across subjects. The concatenated data were then decomposed into 59 spatially independent components (ICs). Repeating this analysis with data from a different resting-state session for the same subjects allowed us to evaluate the test-retest reproducibility of every component. For example, Fig. 1.B displays three typical ICs that were spatially consistent (or correlated) between the two repeated sessions. By pairing the ICs across sessions to maximize the sum of the absolute pairwise correlation coefficients, we identified 50 unique pairs of reproducible components, which showed higher spatial correlations within pairs ($|r|>0.4$, mean±S.D = 0.78±0.14) than across pairs (mean±S.D = 0.04±0.05) (Fig. 1.C). In addition, we found that IC51 and IC55 yielded high spatial correlations with IC36 (r = 0.578) and IC41 (r = 0.63), respectively. But they were not paired, because IC36 and IC41 were better paired with other components while the pairing algorithm did not allow any duplication. We included IC51 and IC55 as reproducible components in subsequent analyses. These results suggest that the ICA-derived spontaneous activity patterns are robust and reproducible in a sub-system spatial scale.

Fig. 2 shows the spatial maps of all 52 reproducible components in a descending order of their test-retest reproducibility. Among them, 92% showed focal patterns, and fewer ICs were distributed (IC#: 34, 35, 49, 50); 30% showed bilateral distributions (IC#: 4, 8, 17, 19, 23, 28, 30, 31, 33, 34, 35, 36, 46, 49, 50); 36% showed clearly anti-correlated patterns with well localized positivity and negativity (IC#: 2, 3, 5, 6, 7, 8, 13, 15, 17, 21, 27, 29, 30, 31, 34, 42, 49, 50). Some of these components not only aligned with existing anatomical borders (e.g. IC36 and IC50 aligned with V1/V2 border), but also aligned with regions with known functional properties (e.g. IC41 and IC55 both matched well with the region MT).

Next, we attempted to identify the components that might be susceptible to artifacts related to head motion. We compared the ICs obtained without and with motion correction (i.e. regressing out head motion correction parameters from voxel time series). Among all the components shown in Fig. 2, three ICs (IC34, IC49, IC50) with distributed patterns did not match to any of the ICs ($|r| < 0.15$) obtained after motion correction (Supplementary Fig. 1A). Thus we attributed these ICs to head motion, and further excluded them for subsequent analyses. All other 49 ICs in Fig. 2 were one-to-one matched to the ICs after head motion correction, showing high spatial correlations for all matched pairs (Supplementary Fig. 1.B). In the following sections, we further segregated and interpreted these 49 ICs by comparing them to existing visual areas or networks.

*Comparing discrete ICA components with existing visual areas*

Since the ICA-derived activity patterns mostly showed discrete regions with well-defined borders (Fig. 2), we further compared such discrete ICs with the visual areas defined with a recently published multi-modal parcellation (MMP) (Glasser et al., 2016). For the primary visual area (V1), four components were found to be sharply confined to V1 (Fig. 3.A). IC8 matched the bilateral foveal representations in V1; IC36 also showed bilateral distributions and corresponded to more peripheral representations; IC26 and IC38 showed unilateral distributions, corresponding to the most peripheral part of the right and left visual fields, respectively; these components did not overlap each other and all aligned with the V1 border. Therefore, V1 consists of multiple intrinsic functional sub-divisions apparently organized according to eccentricity representations, being largely symmetric not only between the left and right hemispheres, but also between the upper and lower sides of the calcarine sulcus. Unlike V1, V2 or V3 did not confine any component within itself. Instead, multiple components spanned across V2 and V3 along either the dorsal (IC18, IC30) or ventral (IC15 and IC9) direction (Fig. 3.B). Compared to those components within V1, the V2/V3 components were less bilaterally symmetric; none of them included regions in both dorsal and ventral pathways.

Beyond those in early visual areas (V1/V2/V3), other discrete components were all anatomically split by the dorsal-ventral division. In the dorsal pathway, some components matched well with existing visual areas (Fig. 4A), including the middle temporal (MT: IC41 on the right hemisphere, IC55 one the left hemisphere), caudal area of inferior parietal cortex (PGp: IC19), dorsal visual transitional area (DVT: IC51 on the right hemisphere, IC39 on the left hemisphere) and parieto-occipital sulcus area 2(POS2: IC4). Some other components were distributed across multiple visual areas, including the third visual area and the area intraparietal 0 (V3B/IP0: IC44 on the right hemisphere, IC20 on the left hemisphere), third visual areas and the fourth visual area (V3A/V3B/V3CD/V4: IC31), the dorsal visual transitional area and the sixth visual area (DVT/V6A: IC28), and the third visual area, the sixth visual area and the seventh visual area (V3A/V6A/V7: IC23); about half of these dorsal components were bilaterally symmetric. Along the ventral pathway, the majority of the components covered multiple visual areas (Fig. 4B), including ventral-medial visual areas (VMV1/VMV2/VMV3: IC42 on the right hemisphere, IC22 on the left hemisphere), para-hippocampal areas (PHA2/PHA3: IC10 on the right hemisphere, IC16 on the left hemisphere), fourth visual areas and the area lateral occipital 2 (V4t/LO2: IC33), the ventral visual complex and the fusiform face complex (VVC/FFC: IC40) and VMV2/PHA2/PHA3 (IC37), except for VMV1 (IC45); about two third of these ventral components were bilaterally symmetric; the rest of them were unilateral.

*Functional modularity in the visual cortex*

For all discrete components, we further evaluated their modular organization based on their temporal correlations. The between-component correlation matrix was calculated for every subject, and then averaged across subjects. The resulting group-level correlation matrix was re-organized into four functional modules based on the Louvain modularity analysis (Blondel et al., 2008; Rubinov and Sporns, 2010). Components within the same module were strongly and positively correlated, whereas components from different modules were weakly or negatively correlated (Fig. 5B, left). Visualizing the distributions of these functional modules on the cortical surface revealed their anatomical segregation (Fig. 5A). The first module included components, each of which was represented by one or two spheres for unilateral (or bilateral) components, over the foveal representations of early visual areas (V1, V2, and V3), whereas the second module was mostly distributed over the peripheral representations of early visual areas. The third module was distributed along the dorsal pathway, and the fourth module was distributed along the ventral pathway (Fig. 5.A). The modularity (Q=0.4966) was statistically significant ($p < 0.0001$, non-parametric permutation test). The between-component correlations within every module were consistently high for most of the subjects, yielding high t statistics, especially for the module over the foveal early visual areas and the module over the ventral visual areas. The t statistics corresponding to the correlations across different modules were generally low and not significant. These results lend support to the notion that intrinsic networks within the visual cortex are organized into functional modules: the dorsal pathway, the ventral pathway, as well as the foveal and peripheral parts of the early visual areas.

*Effects of cortical distance on functional connectivity and modularity*

The observation that the functional modules were anatomically clustered led us to ask whether the functional relationships between individual ICs were entirely attributable to their cortical distances. To address this question, we tried to model the temporal correlations between components as a function of the cortical-surface distance between component centroids, separately evaluated for each hemisphere. The scatter-plots of correlation (r) vs. distance (d) revealed reciprocal relationships, which were modeled as r=1.432/d and r=1.521/d for left and right hemispheres, respectively, based on least-squares estimation (Fig. 6, left). It suggests that fine-scale functional connectivity within a cortical hemisphere depends, at least in part, on cortical distance, in line with findings from previous studies (Dawson et al., 2016; Genc et al., 2016).

Despite this relationship, we further asked whether the modular organization based on functional connectivity (Fig. 5) could be a result of spatial affinity, described as the reciprocal of the distance along the cortical surface. The matrix of spatial affinity was computed for each hemisphere, and reorganized into modules using the same modularity analysis as used for functional connectivity (Fig. 6, right). The left hemisphere contained two anatomical modules: one covering the lateral cortical surface, and the other covering the medial surface. The right hemisphere exhibited similar anatomical modules as the left hemisphere, except that the medial cortical surface was sub-divided into two modules in the right hemisphere. The modularity (Q) of the spatial affinity matrices was 0.2039 and 0.2053, which were statistically significant ($p<0.01$) but notably lower than the modularity of the functional connectivity (Q=0.4966). More importantly, the functional and anatomical modules, on the basis of temporal correlations and spatial affinity respectively, showed different spatial distributions, as comparatively shown in Fig. 5 and Fig. 6. Thus, spontaneous functional connectivity in the visual cortex revealed a stronger and different modular organization and distribution than what were attainable based on spatial affinity alone.

*Functional parcellation of the visual cortex*

Following ICA, we applied the k-means clustering to the ICA weighting vectors at individual locations within the visual cortex, yielding an automated intrinsic parcellation of the visual cortex with a varying level of granularity with the number of clusters (k) being 10, 20, 30, 40. We found that as the number of clusters increased, coarser parcels were progressively sub-divided into finer parcels (Fig. 7). We settled at k=40, which roughly matched the expected number of visual areas (Glasser et al., 2016), and generated a set of well-defined and bilaterally symmetric parcels (Fig. 8). This parcellation based on spontaneous activity was further compared against existing parcellations of the visual cortex, based on the whole-brain multimodal images (i.e. MMP) (Glasser et al., 2016), visual-field maps (Abdollahi et al., 2014), cortical folding patterns (Destrieux et al., 2010), cortical cytoarchitecture (Eickhoff et al., 2005), and cortical myelination (Glasser et al., 2014). It was found that none of the existing parcellations precisely agreed with the fully-automated and data-driven parcellation reported here. In particular, our parcellation did not match with those based on cortical retinotopy, cytoarchitecture, and folding, but matched relatively better with the cortical myelination and the MMP. In our parcellation, the outer contour of cortical parcels that covered early visual areas tended to align well with the steep gradients of cortical myelination. Our reported cortical parcels that covered the high-level visual areas tended to align reasonably well with the corresponding parcels in MMP; the alignment was not one to one. Compared to MMP, our parcellation was coarser in high-level visual areas, but finer in low-level visual areas.

*Parcellation in the level of single subjects*

In addition to the group-level analysis, we also applied dual regression to the dataset, in an attempt to obtain the corresponding ICA patterns from individual subjects. The individual-level ICA patterns were generally consistent with the group-level ICA patterns and comparable across subjects (Fig. 9.A). On the basis of individual-specific ICs, the visual cortex could be parcellated (k = 40) for each subject, yielding cortical parcels apparently noisy but generally similar to the group-level parcellation (Fig. 9.B).

**Discussion**

We characterized the network patterns emerging from spontaneous resting-state activity within the human visual cortex. On the basis of such patterns and their interactions, we delineated the intrinsic functional parcellation and organization of the visual cortex. Here we report that fine-scale intrinsic visual cortical networks are not organized by the patterns of cortical retinotopy, folding, or cytoarchitecture, but align with the gradient of cortical myelination, and are segregated into functional modules specific to the ventral and dorsal visual streams.

*Whole-brain vs. fine-scale functional networks*

The majority of resting-state fMRI literature focuses on intrinsic functional networks in the whole brain scale (Smith et al., 2013). Large-scale networks are supported by long-range structural connections (Yeo et al., 2011), and reflect coarsely defined functions (Smith et al., 2009). However, in a finer scale within the visual cortex, bi-directional structural connections co-exist over short distances (Felleman and Van Essen, 1991; Salin and Bullier, 1995), forming the network basis of vision (Rao and Ballard, 1999). The patterns and dynamics of fine-scale intrinsic networks may indicate how visual information is coded in spontaneous brain activity (Kenet et al., 2003), and offer a more specific clue on the functional role of spontaneous activity in shaping perception or behavior (Wilf et al., 2015).

To explore the topographic organization of fine-scale visual networks, it is necessary to confine the analysis to the functional connectivity profile within the visual system. Otherwise, in the whole-brain scale, the connectivity profile between a seed location in the visual cortex and the rest of the brain, as a function of locations, includes mostly the remote locations that are not or non-specifically associated with the seed location. As a result, the difference in the whole-brain connectivity profile of two distinct seed locations becomes subtle, even if they may actually interact with different sets of vison-related brain locations. Focusing on a fine scale improves the sensitivity to differentiate the topographic difference in functional connectivity of specific interest to vision. In line with this notion, previous studies have shown that fine-scale functional networks may be obscured by large-scale network activity (Raemaekers et al., 2014), and thus appear to exhibit a coarse topographic organization (Nir et al., 2006; Yeo et al., 2011).

Within the visual cortex, we used data-driven ICA to explore the multivariate voxel patterns and dynamics, instead of bivariate correlations between voxels or areas, as in previous studies (Heinzle et al., 2011; Yeo et al., 2011; Butt et al., 2013; de Zwart et al., 2013; Gravel et al., 2014; Raemaekers et al., 2014; Arcaro et al., 2015; Bock et al., 2015; Striem-Amit et al., 2015; Wilf et al., 2015; Dawson et al., 2016; Genc et al., 2016). In the whole-brain scale, ICA and seed-based correlation analyses have been shown to reveal similar network patterns (Van Dijk et al., 2010). However, structural connections are much denser at a reduced spatial scale or distance (Bassett and Bullmore, 2006), giving rise to more complex patterns of functional interactions. This makes ICA a more preferable method for fine-scale network mapping.

*Fine-scale visual networks are not retinotopically organized*

Our data suggest that resting state networks within the visual cortex are robust in group (Fig. 1) and individual (Fig. 9) levels, and well-organized in space (Figs. 2 through 4) and time (Figs. 5, 6). Here, the results obtained with data-driven ICA extend the previous findings obtained by correlational analyses of rs-fMRI signals in the visual cortex.

However, unlike some prior studies (Heinzle et al., 2011; Gravel et al., 2014; Raemaekers et al., 2014), we did not find any evidence for the retinotopic organization of resting state activity beyond early visual areas. Perhaps, the exception was only in V1, where activity patterns were found to agree with eccentricity representations (Fig. 3A), consistent with previous findings (Yeo et al., 2011; Arcaro et al., 2015; Wilf et al., 2015). Beyond V1, spontaneous activity patterns were independent of either the eccentricity or the polar angle (Fig. 3). Even in V1, resting state activity was correlated between the left and right hemispheres (Fig. 3), although the two hemispheres correspond to different hemi-fields in the visual space. Note that the left and right V1 areas have little or no callosal connections (Tootell et al., 1998) to directly support their synchronization. The inter-hemispherical V1 correlation is most likely due to a common input to both hemispheres. The topography of this common input seems retinotopically non-specific, at least in terms of the polar angle.

Arguably, the eccentricity-dependent intrinsic activity patterns in V1 may be coincidental, and reflect the relative distributions of magnocellular (M) and parvocellular (P) projections that happen to vary with eccentricity. Previous studies have shown that P cells, relative to M cells, over-represent the central vision but under-represent the periphery in lateral geniculate nuclei (LGN) (Connolly and Van Essen, 1984; Schiller et al., 1990) and V1 (Baseler and Sutter, 1997; Azzopardi et al., 1999). The ratio between P and M projections to/from V1 notably decreases with eccentricity (Baseler and Sutter, 1997). Note that P and M pathways convey distinct visual attributes, but share the same retinotopic maps (Nassi and Callaway, 2009; Denison et al., 2014). Although the representation of the M-to-P ratio seems similar as the eccentricity representation in V1, the M-P pathways bear a different organization specific to visual streams as opposed to visual locations. Caution should be exercised before interpreting an eccentricity-dependent pattern alone as evidence for the retinotopic organization.

*Resting-state activity reflects feedback visual-network interactions*

We further speculate that the common input to V1, which drives visual-stream specific resting-state activity, arises from top-down modulations through feedback connections. In the visual hierarchy (Felleman and Van Essen, 1991), the population receptive field becomes larger and less specific from lower to higher visual areas (Wandell et al., 2007), to progressively converge information in the visual space through feedforward connections. While the feedforward connections are retinotopically organized, the feedback corticocortical connections are not so (Salin and Bullier, 1995). Through feedback, the top-down modulations transfer information about a large or even the whole visual field to cortical locations with specific receptive fields (Salin and Bullier, 1995), driving network activity patterns away from being retinotopically specific. It is conceivable that the organization of feedback connections plays a defining role to resting state networks within the visual cortex. This is because when feedforward pathways are not driven by fluctuating external inputs, feedback pathways are modulated by the brain's intrinsic activity.

In addition, feedback connections in the visual system are generally separated by the ventral and dorsal pathways (Salin and Bullier, 1995; Gilbert and Li, 2013), functionally specialized for recognition and action, respectively (Goodale and Milner, 1992; Ungerleider and Haxby, 1994). Such a visual-stream-specific organization also applies to feedback connections from V1 to LGN (Briggs and Usrey, 2009). It lends support for our interpretation that feedback projections serve the structural network basis of the observed modular organization of intrinsic fine-scale functional networks distributed along the ventral and dorsal pathways (Fig. 5). As the ventral and dorsal pathways become intricate in early visual areas, fine-scale network patterns in V1/V2/V3 constitute the third functional module (Fig. 5).

*Intrinsic functional parcellation of the visual cortex*

On the basis of fine-scale resting-state activity patterns, the functional parcellation of the visual cortex appeared to be notably different from other parcellations based on cortical folding, retinotopy, and cytoarchitecture. The discrepancy with the cortical folding is perhaps reasonable, because the relationship between cortical morphology and functional organization is elusive and indirect, despite a developmental linkage likely between them as proposed elsewhere (Benson et al., 2012; Ronan and Fletcher, 2015). The discrepancy with visual field maps is perhaps also understandable, for the reasons elaborated in previous sections. In addition, the parcellation based on functional connectivity is applicable to all locations in the visual cortex, whereas the visual field is not mapped at all visual areas. For the apparently different parcellations based on regional cellular composition and inter-regional functional connectivity suggests a lack of one-to-one relationships between cell types and functional networks.

Our results show that the functional networks and parcels seem to align with the gradient of myelination (Fig. 8). We speculate that spontaneous activity shapes myelin density. It has been shown that electrical activity may promote myelination(Gibson et al., 2014), and that functional organization and cortical myelination may co-vary given plasticity (Fields, 2015; Hunt et al., 2016). As such, greater myelin density may imply greater functional specificity but less plasticity: early sensory areas are more functionally specific with greater myelination or less plasticity, whereas higher-order or multi-sensory areas are less functionally specific with less myelin density and less plasticity (Glasser et al., 2014).

**Acknowledgments**

This work was supported in part by NIH R01MH104402 (Z Liu) and Purdue University. Authors have no conflict of interest.

**Figure Caption**

### A. Visual Cortical Mask

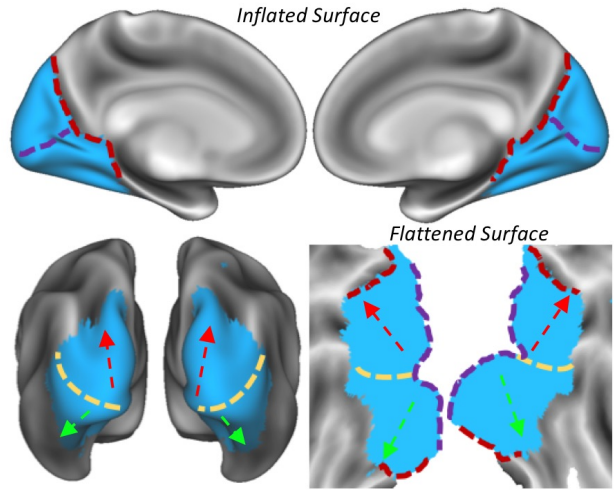

### B. Examples of Reproducible ICs

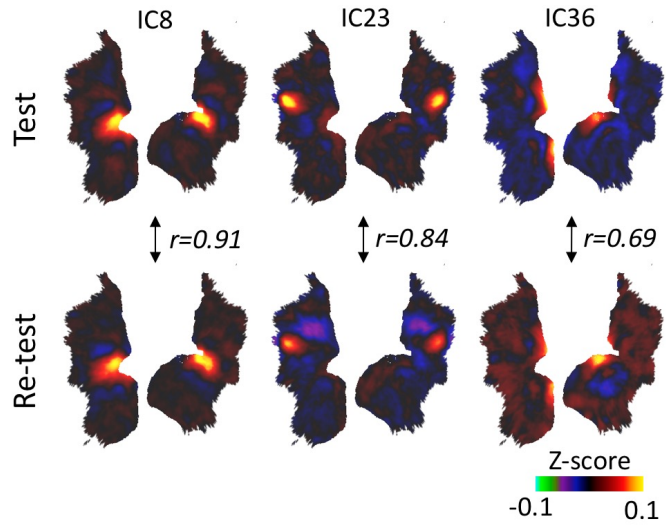

### C. Test-retest Reproducibility

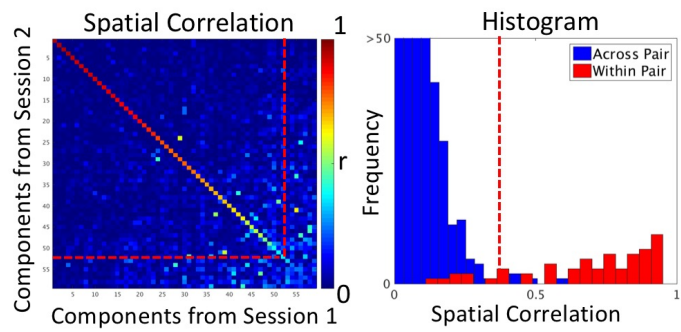

**Figure 1. Visual cortical mask and reproducibility of fine-scale network patterns. A.** The visual cortex mask (in blue) is illustrated on inflated and flattened cortical surfaces. Reference lines mark the occipital-parietal sulcus (red dash line), the calcarine sulcus (purple dash line), a rough ventral-dorsal division (yellow dash line), from where the ventral and dorsal pathways are along the green and red arrows. **B.** Three examples of reproducible ICs that exhibit high correlations (r) in their spatial patterns. The color scale represents the z-score and is ranged from -0.1 to 0.1. **C.** Shown in the left is the spatial correlation matrix between the ICs obtained from two repeated resting-state sessions. The diagonal elements correspond to uniquely paired ICs; the off-diagonal elements are between unpaired ICs. The red line represents a correlation threshold (r=0.4), by which a pair of ICs was considered reproducible between sessions. Shown in the middle is the p-values associated with the spatial correlations. Shown in the right panel is the discrete histograms of the correlations for the paired (red) and unpaired (blue) ICs.

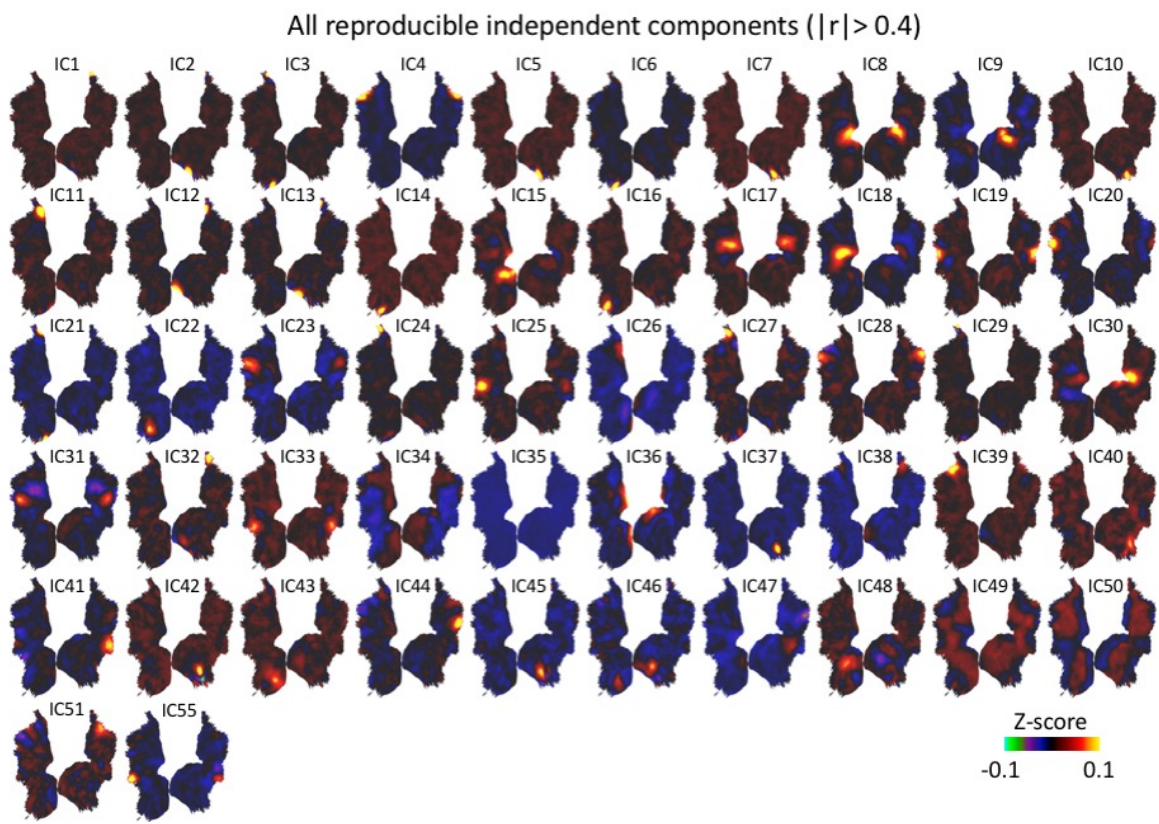

**Figure 2. 52 reproducible independent components with spatial correlation greater than 0.4. A.** The ICs were numbered in a descending order of their test-retest reproducibility. IC51 and IC55 were not
22

optimally matched by our algorithm, but they yielded spatial correlations of 0.58 and 0.63 with IC39 and IC41 respectively. Therefore, they are also considered to be reproducible. **B.** The remaining ICs that have less consistency across the two datasets.

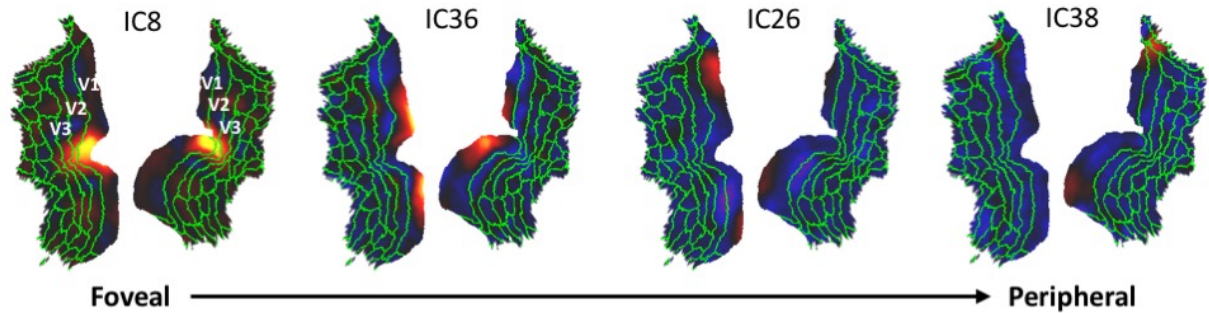

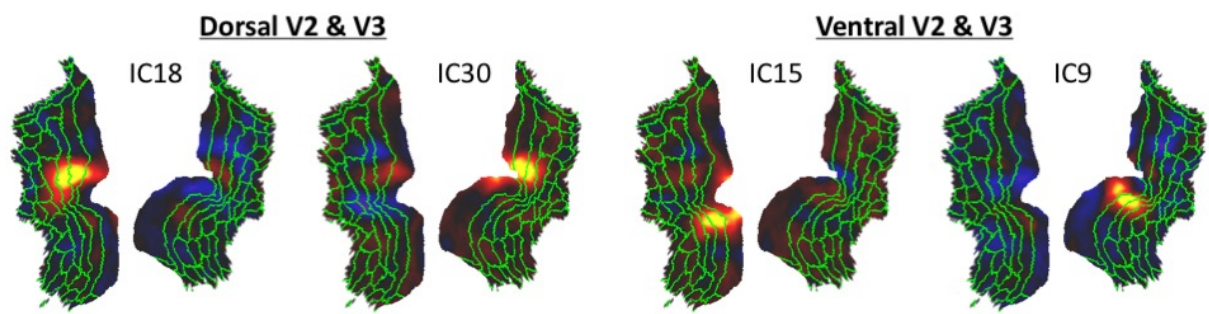

**Figure 3. Discrete ICs within the early visual areas (V1/V2/V3). A.** Four ICs within V1. **B.** Four ICs within V2 and V3. The green lines are the borders of the multi-modal parcellation (Glasser et al., 2016), where V1/V2/V3 are defined.

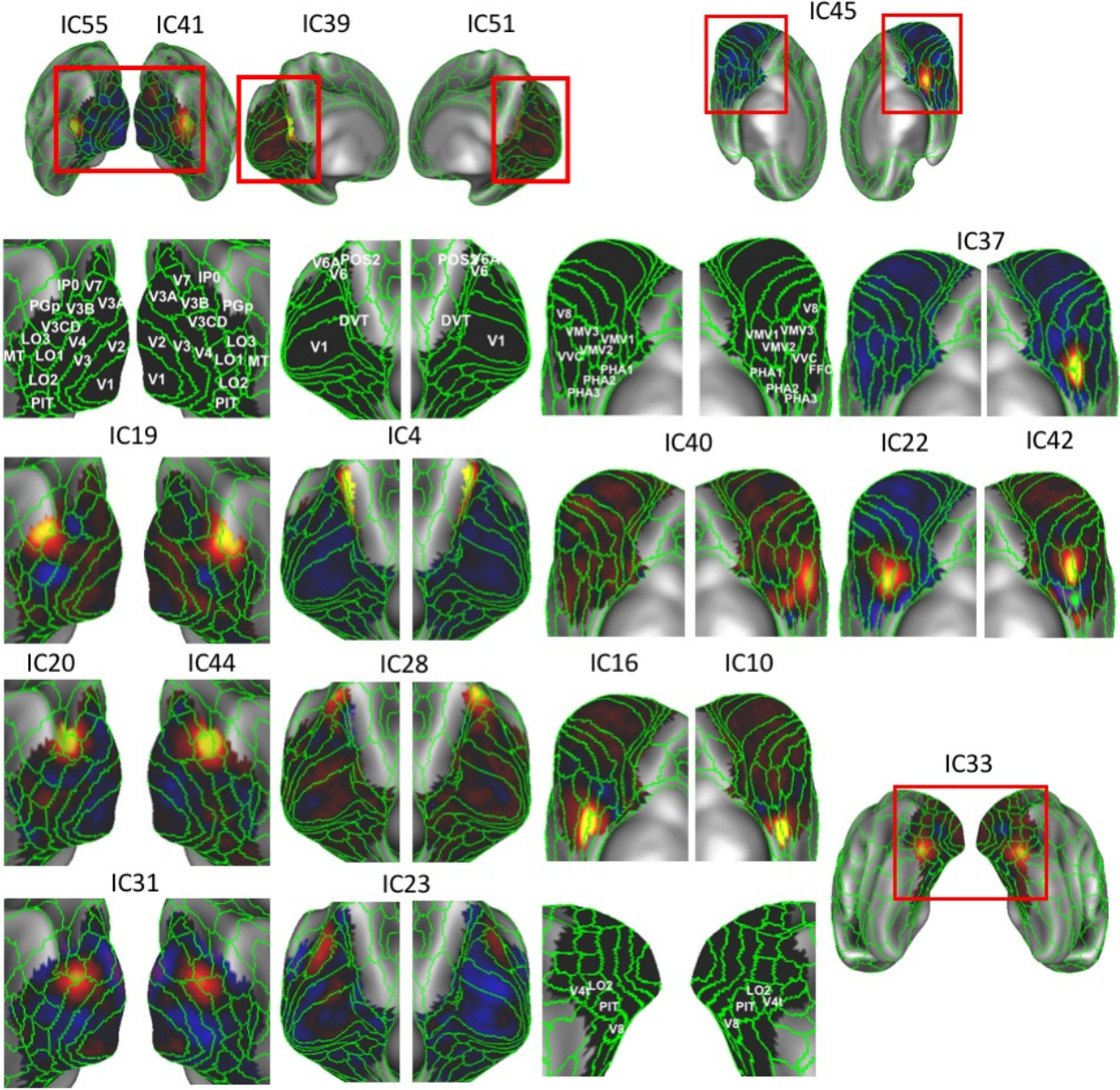

**Figure 4. Discrete ICs along the dorsal pathway and ventral pathway. A.** The left panel shows the ICs that match well with existing lateral visual areas in the dorsal pathway. The right panel shows the ICs that match well with existing medial visual areas in the dorsal pathway. **B.** Example ICs that match well with existing medial visual areas in the ventral pathway such as VMV1-3, FFC and VVC. The green lines mark the existing visual areal borders according to the MMP.

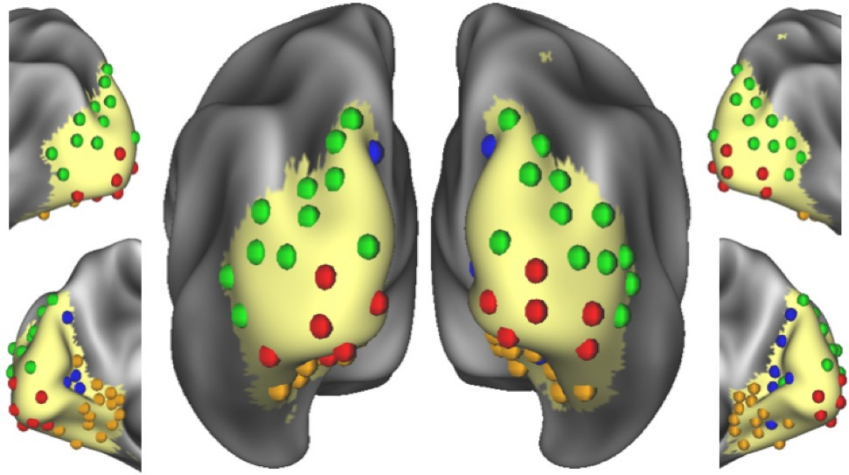

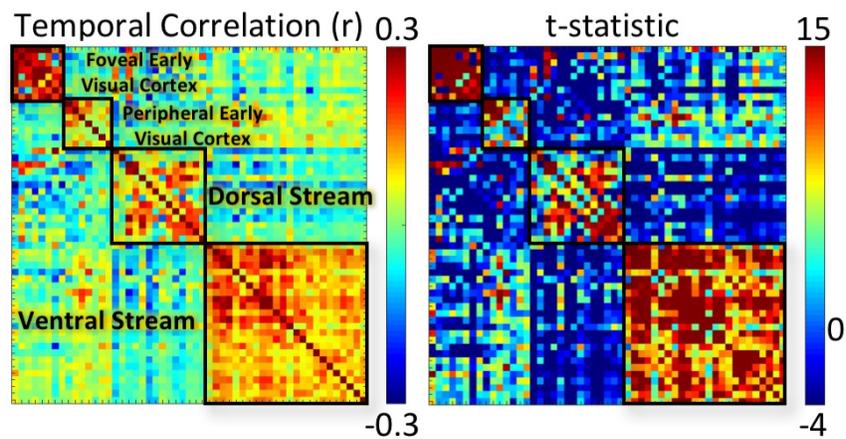

**Figure 5. Functional modules in the visual cortex. A.** Each component is represented with a sphere, colored coded by its modular membership, and placed at the peak location of its spatial map. **B.** The matrix on the left shows the temporal correlations (r) between ICs, organized into four modules (in (1) foveal early visual cortex, (2) peripheral visual cortex, (3) dorsal pathway, (4) ventral pathway). The matrix on the right shows the t-statistics associated with the temporal correlations (z) between ICs, organized in the same order as the matrix on the left.

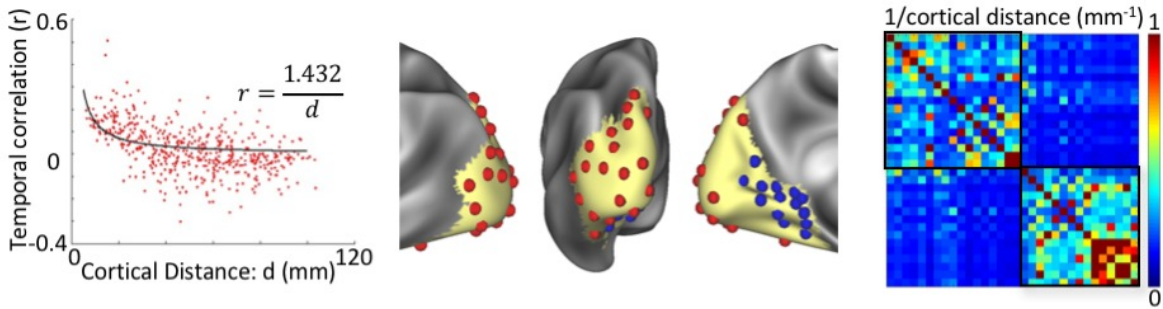

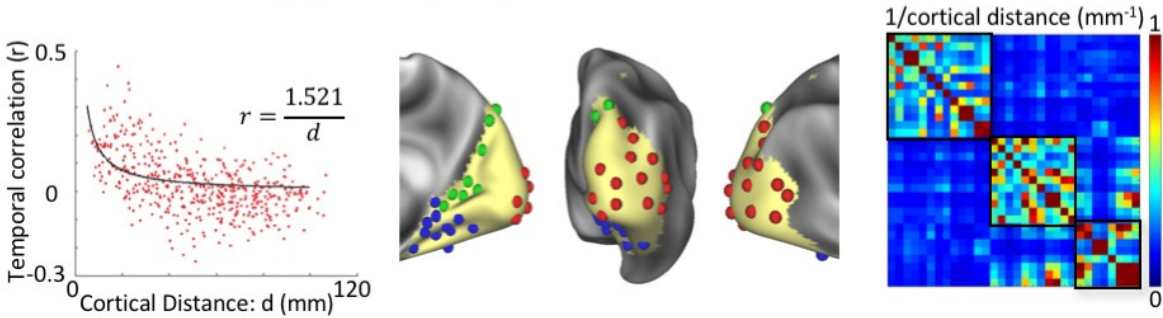

**Figure 6. The effect of cortical distance on functional connectivity and modularity. A.** Left panel: The scatter-plots of correlation (r) vs. distance (d), which were modeled as r=1.432/d based on least-squares estimation, in the left hemisphere. Right panel: The matrix shows the spatial-affinity between component centroids within the left hemisphere. Middle panel: Each component is represented with a sphere, colored coded by its modular membership, and placed at the peak location of its spatial map. **B.** Left panel: The scatter-plots of correlation (r) vs. distance (d), which were modeled as r=1.521/d based on least-squares estimation, in the right hemisphere. Right panel: The matrix shows the spatial-affinity between component centroids within the right hemisphere. Middle panel: Each component is represented with a sphere, colored coded by its modular membership, and placed at the peak location of its spatial map.

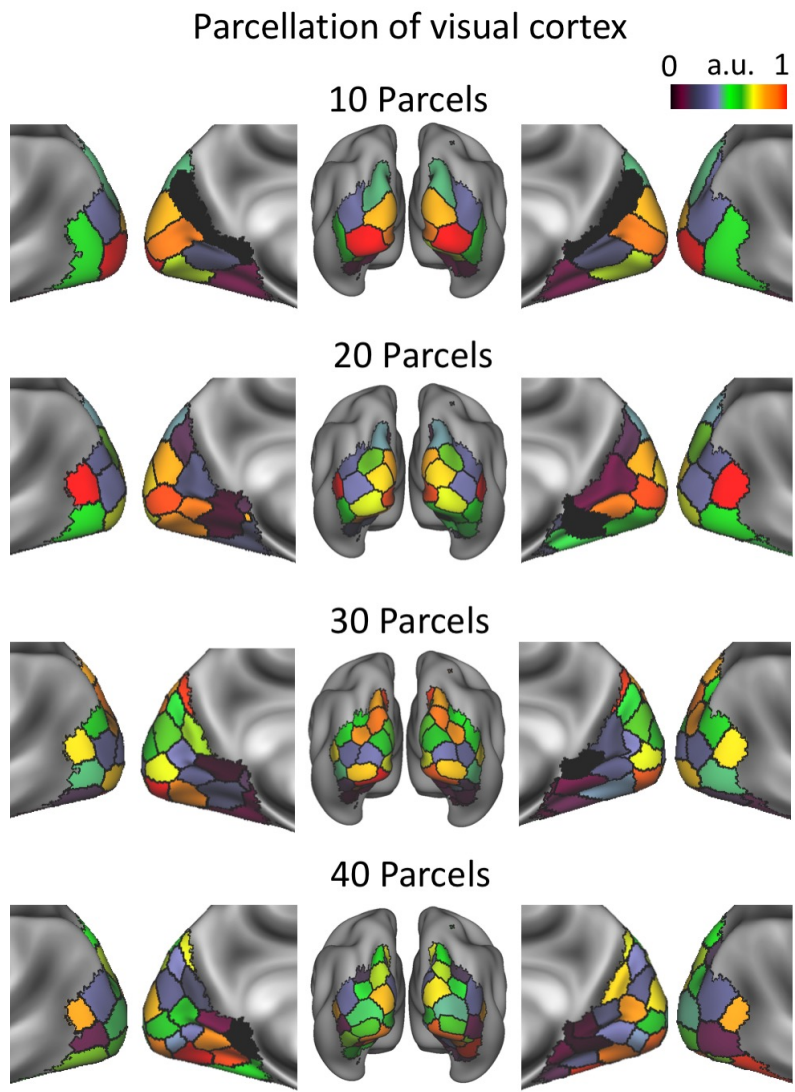

**Figure 7. Functional parcellation of the visual cortex with varying number of parcels (k=10, 20, 30, 40).** The parcels are color-coded from 0 to 1, sorted in an ascending order according to the averaged sum of distances within each parcel.

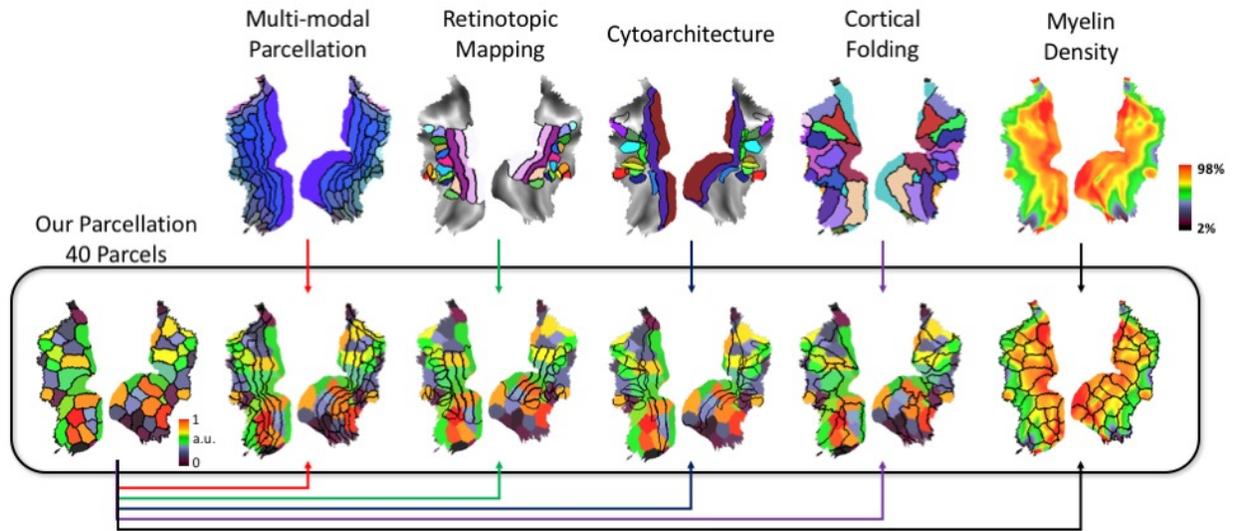

**Figure 8. Comparing our parcellation (k = 40) with existing visual cortex parcellations.** Our functional parcellation is compared against (1) whole-brain multimodal images (i.e. MMP) (Glasser et al., 2016); (2) visual-field maps (Abdollahi et al., 2014); (3) visual-field maps (Abdollahi et al., 2014); (4) cortical folding patterns (Destrieux et al., 2010); (5) cortical myelination (Glasser et al., 2014).

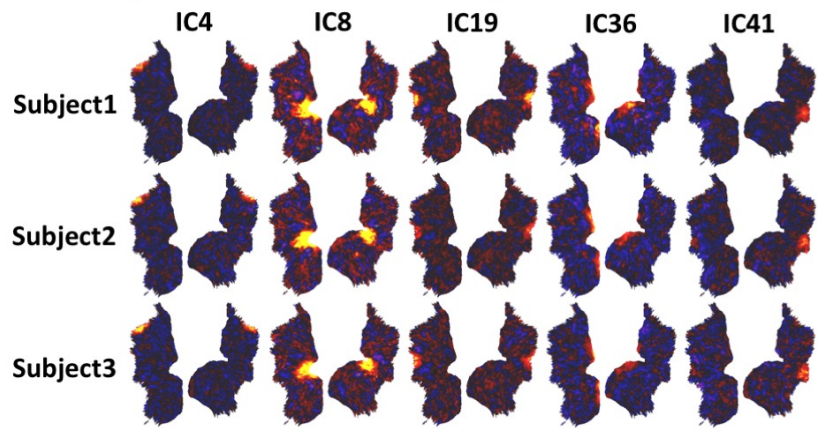

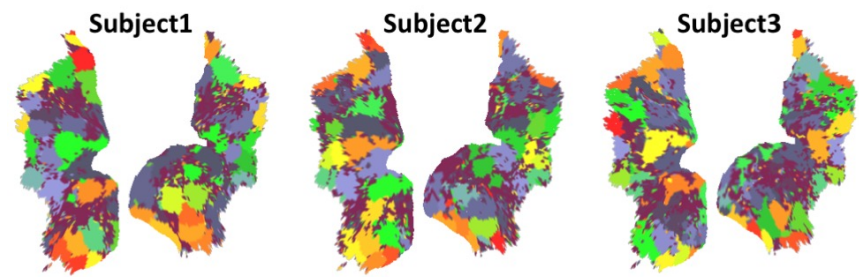

**Figure 9. Independent components and parcellations from three individual subjects obtained through dual regression.** **A.** Five example ICs from three subjects. **B.** The individual-level parcellations obtained from three example subjects.

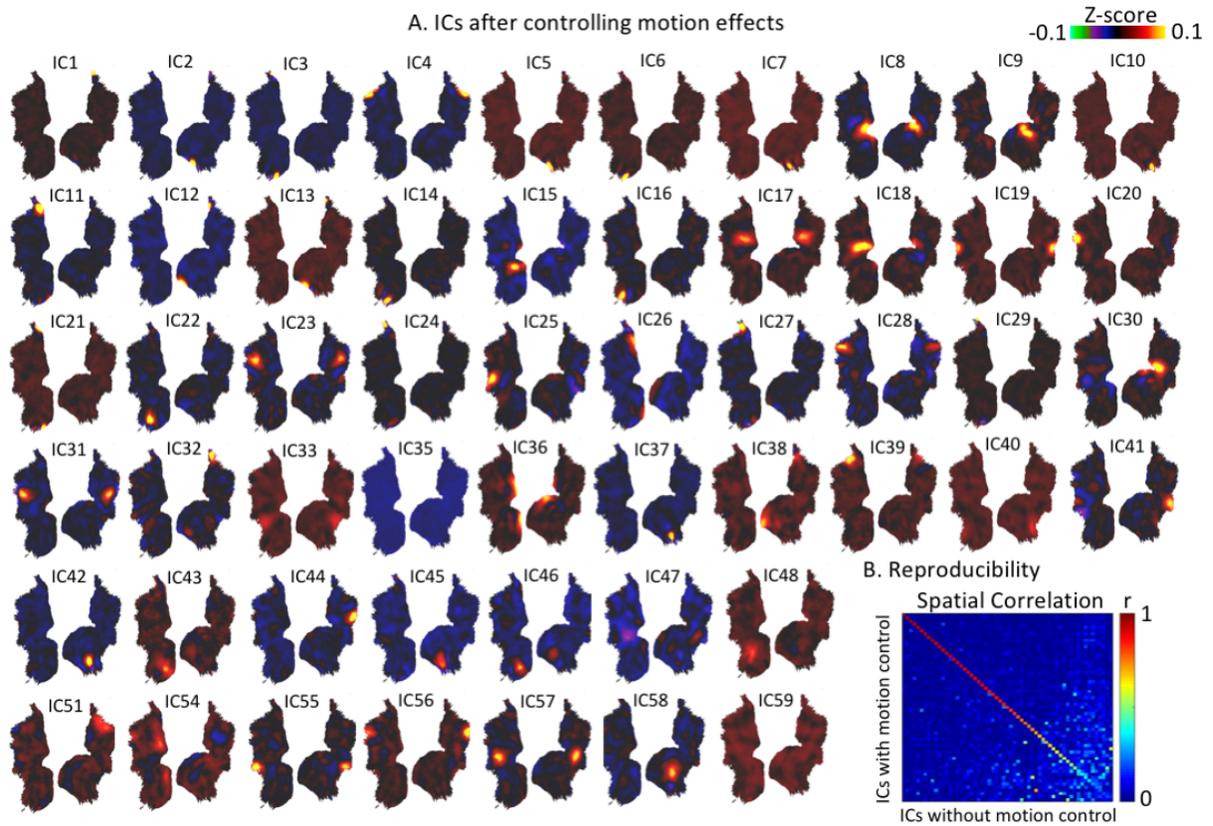

**Supplementary Figure 1. 54 independent components obtained from motion-controlled fMRI dataset with spatial correlation greater than 0.15.** **A.** The 54 ICs are spatially consistent with those observed in Fig. 2. **B.** Spatial cross correlation between ICs with and without controlling for motion effects.